\documentstyle[pra,aps,epsf]{revtex}

\newcommand{\beq}{\begin{equation}}
\newcommand{\eeq}{\end{equation}}
\newcommand{\bes}{\begin{eqnarray}}
\newcommand{\ees}{\end{eqnarray}}

\begin{document}
\draft
\thispagestyle{empty}
\title{Lateral projection as a possible explanation of the
nontrivial boundary dependence of the Casimir force}

\author{
 G.~L.~Klimchitskaya\footnote{On leave from
North-West Polytechnical Institute,St.Petersburg, Russia.  
Electronic address:  galinak@fisica.ufpb.br},
S.~I.~Zanette, and A.~O.~Caride
}

\address
{
Centro Brasileiro de Pesquisas F\'{\i}sicas, Rua Dr.~Xavier Sigaud, 150\\
Urca 22290--180, Rio de Janeiro, RJ --- Brazil
}

\maketitle
\begin{abstract}
We find the lateral projection of the Casimir force for a configuration
of a sphere above a corrugated plate. This force tends to change the
sphere position in the direction of a nearest corrugation maximum.
The probability distribution describing different positions of a sphere
above a corrugated plate is suggested which is fitted well with
experimental data demonstrating the nontrivial boundary dependence
of the Casimir force.
\end{abstract}

%%%%%%%%%%%%

Considerable recent attention has been focused on the Casimir effect
\cite{1,2}. On the theoretical side much work was done to investigate
the corrections to the Casimir force due to the finite conductivity
of the boundary metal \cite{3,4,5,6}, nonzero temperature
\cite{7,8,9}, and surface roughness \cite{10,11,11a}. In the 
experimental field the new precision measurements of the Casimir force
between metallic surfaces of a plane disk and a spherical lens (or
a sphere) were performed \cite{12,13,14,15}. The experimental
results correlate well with the theoretical expressions taking into account
all the corrections mentioned above. This provided a way to obtain
new stronger constraints on the constants of Yukawa-type
additional terms to Newtonian gravitational law predicted by the
unified gauge theories, supersymmetry, supergravity, and string
theory \cite{16,17,18,19,20}. The Casimir effect has assumed a new
meaning as a tool for investigation of fundamental interactions and
their unification. Because of this, the detailed analyses of the
fit of the theory to the data takes on great significance.

In Ref.~\cite{21} the Casimir force between an aluminum coated plate
with sinusoidal corrugations and a large sphere was measured using an 
atomic force microscope. It was concluded that the measured force shows 
significant deviation from the perturbative theory which takes into
account the periodic corrugation of the plate in the surface
separation. In the absence of corrugations the same theory shows good
agreement with the measured Casimir force. These together were considered
in \cite{21} to represent 
the nontrivial boundary dependence of the Casimir force and
until recently has no theoretical explanation (in line with \cite{21}
dependence of this kind is to be expected due to diffractive effects
associated with corrugated surface).

Here we present the perturbative calculation for both vertical and lateral 
Casimir force acting in the configuration of a sphere situated above 
a corrugated 
plate (note that the lateral force arises due to the absence of
translational symmetry on a plate with corrugations). Our study revealed
that the lateral force acts upon the sphere in such a way that it tends 
to change its position in the direction of a nearest maximum point
of the vertical Casimir force (which coincides with the maximum point of 
corrugations). In consequence of this, the assumption made in \cite{21} that 
the locations of the sphere above different points of a corrugated surface 
are equally probable can be violated. As indicated below, the diverse 
assumptions on the probability distribution describing location of the
sphere above different points of the plate result in an essential change
in force-distance relation. In such a manner the perturbation theory
taking the lateral force into account may work for a case of corrugated
plate. Notice that the fundamental importance of the lateral Casimir
force acting between the corrugated boundaries was discussed in
\cite{22,23,24}.

We start with the configuration of polystyrene sphere above a 
$7.5\times 7.5\,\mbox{mm}^2$ plate with periodic uniaxial sinusoidal
corrugations. Both the sphere and the plate were coated with 250\,nm of
$Al$, and 8\,nm layer of $Au/Pd$. For the outer $Au/Pd$ layer 
transparencies greater than 90\% were measured at characteristic frequences 
contributing into the Casimir force. The diameter of the sphere was
$2R=(194.6\pm 0.5)\,\mu$m. The surface of the corrugated plate is
described by the function
\beq
z_s(x,y)=A\sin\frac{2\pi x}{L},
\label{1}
\eeq
\noindent
where the amplitude of the corrugation is $A=(59.4\pm 2.5)\,$nm and
its period is $L=1.1\,\mu$m. The mean amplitude of the stochastic
roughness on the corrugated plate was $A_p=4.7\,$nm, and on the
sphere bottom --- $A_s=5\,$nm.

As is seen from Eq.~(\ref{1}) the origin of the $z$-axis is taken such 
that the mean value of the corrugation is zero. We represent by $a$ the
separation between the zero corrugation level and the sphere bottom
as they are not taking into account the surface roughness. 
The minimal value of $a$ is determined, approximatelly, by
$a_0\approx A+A_p+A_s+2h\approx 130\,$nm, where $h\approx 30\,$nm is 
the height of the highest occasional rare $Al$ crystals which prevent
the intimate contact between the sphere bottom and the maximum point
of the corrugation. The independent measurement of $a_0$ by use of
electrostatic force results in $a_0=(132\pm 5)\,$nm \cite{21}. For the
case of the Casimir force measurements, i.e. with the addition of two
transparent $Au/Pd$ layers, this leads to $a_0=(148\pm 5)\,$nm.
We consider here the measurement data in the interval
$164.5\,\mbox{nm}\leq a\leq 400\,$nm (for larger $a$ the Casimir force 
is less than the experimental uncertainty).

Let us take up first the vertical Casimir force acting between a corrugated
plate and a sphere. A sinusoidal corrugation of Eq.~(\ref{1}) leads to
the modification of the Casimir force between a flat plate and a sphere. 
The modified force can be calculated by the
averaging over the period 
\beq
F(a)=\int\limits_{0}^{L}dx\rho(x)\,F_{ps}\left(d(a,x)\right).
\label{2}
\eeq
\noindent
Here $d(a,x)$ is the separation between the sphere bottom and
the point $x$ on the surface of corrugated plate
\beq
d(a,x)=a-A_p-A_s-A\sin\frac{2\pi x}{L}.
\label{3}
\eeq
\noindent
$F_{ps}$ is the Casimir force acting between a flat plate and a sphere with
account of corrections due to finite conductivity of the boundary
metal \cite{6}
\beq
F_{ps}(d)=F_{ps}^{(0)}(d)
\sum\limits_{i=0}^{4} c_i\left(\frac{\delta_0}{d}\right)^i,
\label{4}
\eeq
\noindent
where $F_{ps}^{(0)}(d)=-\pi^3R\hbar c/(360 a^3)$ is the same force
for a perfect metal, and the numerical coefficients are
\beq
c_0=1,\quad c_1=-4,\quad c_2=\frac{72}{5},\quad
c_3=-\frac{320}{7}\left(1-\frac{\pi^2}{210}\right),
\quad
c_4=-\frac{400}{3}\left(1-\frac{163\pi^2}{7350}\right).
\label{5}
\eeq
\noindent
Note that the temperature corrections are neglegible at the
separations under consideration.
The quantity $\delta_0$ from (\ref{4}) is the penetration depth of the 
zero-point oscillations into the metal given by 
$\delta_0=\lambda_p/(2\pi)$, $\lambda_p$ is the plasma wavelength
($\lambda_p\approx 100\,$nm for $Al$). Note that Eq.~(\ref{4}) was obtained
by means of Proximity Force Theorem \cite{25} and perturbation
expansion of the Lifshitz expression for the Casimir energy between 
two plane parallel plates.

The quantity $\rho(x)$ from Eq.~(\ref{2}) describes the probability
distribution of the sphere positions above different points $x$
belonging to one corrugation period. In \cite{21} the uniform
distribution was assumed $\left(\rho(x)=1/L\right)$ which is to say
that the sphere is located above all points $x$ belonging to the
interval $0<x<L$ with equal probability. The right-hand side of
Eq.~(\ref{2}) was expanded in powers of a small parameter
$A/(a-A_p-A_s)$. This expansion had shown significant deviations from
the measured data while Eq.~(\ref{4}) is in excellent agreement with
data for all $d\geq\lambda_p$ \cite{6} in the limit of zero
amplitude of corrugation.

Now we turn to the lateral projection of the Casimir force which was
not considered in \cite{21}. The lateral projection is nonzero only in the
case of nonzero corrugation amplitude. In \cite{21} neither the lateral
force nor the force constant relative to the horizontal displacement of 
a cantilever were measured. Because of this it is not reasonable to
develop the complete theory describing the shifts of a sphere under
the influence of a lateral Casimir force and giving the possibility to
find theoretically the probability distribution $\rho(x)$ from
Eq.~(\ref{2}). In this connection we restrict ourselves to calculation 
of the lateral force in the simplest case of the ideal metal.

This can be achieved by applying the additive summation method of the
retarded interatomic potentials over the volumes of a corrugated plate
and a sphere with subsequent normalization of the interaction constant
\cite{5,11,11a}. Alternatively the same result is obtainable by the
Proximity Force Theorem \cite{25} (for the case of the nonretarded
van der Waals force these approaches were applied, e.g., in \cite{25a}).
Let an atom of a sphere be situated at a point with the coordinates
($x_A,y_A,z_A$) in the coordinate system described above. Integrating
the interatomic potential $V=-C/r_{12}^7$ over the volume of
corrugated plate ($r_{12}$ is a distance between this atom and the atoms
of a plate) and calculating the lateral force projection according to
$-\partial V/\partial x_A$ one obtains \cite{11,11a}
\beq
F_x^{(A)}(x_A,y_A,z_A)=\frac{4\pi^2n_pC}{5z_A^5}\,\frac{A}{z_A}
\frac{z_A}{L}
\left[\cos\frac{2\pi x_A}{L}
+\frac{5}{2}\frac{A}{z_A}
\sin\frac{4\pi x_A}{L}\right],
\label{6}
\eeq
\noindent
where $n_p$ is the atomic density of a corrugated plate. Eq.~(\ref{6}) is
obtained by perturbation expansion of the integral (up to second order)
in small parameter $A/z_A$.

We can represent $x_A=x_0+x$, $y_A=y_0+y$, $z_A=z_0+z$ where 
($x_0,y_0,z_0$) are the coordinates of the sphere bottom in the above
coordinate system, and ($x,y,z$) are the coordinates of the sphere atom
in relation to the sphere bottom. The lateral Casimir force acting
upon a sphere is calculated  by the integration of (\ref{6}) over the
sphere volume and subsequent division by the normalization factor
$K=24Cn_pn_s/(\pi\hbar c)$ obtained by comparison of additive and exact
results for the configuration of two plane parallel plates \cite{5}
\beq
F_x(x_0,y_0,z_0)=\frac{n_s}{K}
\int\limits_{V_s}d^3r\,F_x^{(A)}(x_0+x,y_0+y,z_0+z),
\label{7}
\eeq
\noindent
where $n_s$ is the atomic density of sphere metal.

Let us substitute Eq.~(\ref{6}) into Eq.~(\ref{7})  neglecting the
small contribution of the upper semisphere which is of order
$z_0/R<4\times 10^{-3}$ comparing to unity. In a cylindrical coordinate
system the lateral force acting upon a sphere rearranges to the form
\bes
&&
F_x(x_0,y_0,z_0)=\frac{\pi^3\hbar c}{30}\frac{A}{L}
\left[\cos\frac{2\pi x_0}{L}
\int\limits_{0}^{R}\rho d\rho
\int\limits_{0}^{R-\sqrt{R^2-\rho^2}}
\frac{dz}{(z_0+z)^5}
\int\limits_{0}^{2\pi}
d\varphi\cos\left(\frac{2\pi\rho}{L}\cos\varphi\right)\right.
\nonumber \\
&&\phantom{aaa}
+\frac{5}{2}A\left.
\sin\frac{4\pi x_0}{L}
\int\limits_{0}^{R}\rho d\rho
\int\limits_{0}^{R-\sqrt{R^2-\rho^2}}
\frac{dz}{(z_0+z)^6}
\int\limits_{0}^{2\pi}
d\varphi\cos\left(\frac{4\pi\rho}{L}\cos\varphi\right)\right].
\label{8}
\ees

Using the standard formulas from \cite{26} the integrals with respect to
$\varphi$ and $z$ are taken explicitly. Preserving only the lowest
order terms in small parameter $x_0/R<10^{-2}$ we arrive at
\beq
F_x(x_0,y_0,z_0)=-\frac{\pi^4\hbar c}{60z_0^4}\frac{A}{L}
\left[\cos\frac{2\pi x_0}{L}
\int\limits_{0}^{R}\rho d\rho
J_0\left(\frac{2\pi\rho}{L}\right)
+2\frac{A}{z_0}
\sin\frac{4\pi x_0}{L}
\int\limits_{0}^{R}\rho d\rho
J_0\left(\frac{4\pi\rho}{L}\right)\right],
\label{9}
\eeq
\noindent
where $J_n(z)$ is Bessel function.

Integrating in $\rho$ the final result is obtained
\beq
F_x(x_0,y_0,z_0)=3F_{ps}^{(0)}(z_0)\frac{A}{z_0}
\left[\cos\frac{2\pi x_0}{L}\,J_1\left(\frac{2\pi R}{L}\right)
+\frac{A}{z_0}
\sin\frac{4\pi x_0}{L}\,J_1\left(\frac{4\pi R}{L}\right)\right],
\label{10}
\eeq
\noindent
where the vertical Casimir force $F_{ps}^{(0)}$ for ideal metal was
defined after Eq.~(\ref{4}).

As is seen from Eq.~(\ref{10}) the lateral Casimir force takes zero value 
at the extremum points of the corrugation described by Eq.~(\ref{1}).
The lateral force achieves maximum at the points $x_0=0,\,L/2$ where the
corrugation function is zero. If the sphere is situated to the left of
a point $x_0=L/4$ (maximum of corrugation) it experiences a positive
lateral Casimir force. If it is situated to the right of $x_0=L/4$ the
lateral Casimir force is negative. In both cases the sphere tends to change 
its position in the direction of a corrugation maximum which is the
position of stable equilibrium. The situation here is the same as for
an atom near the wall covered by the large-scale roughness \cite{11}.
That is the reason why the different points of a corrugated plate are
not equivalent and the assumption that the locations of the sphere
above them are described by the uniform probability distribution
may be unjustified.

On this basis, one may suppose that the probability distribution under
consideration is given by
\beq
\rho(x)=\left\{
\begin{array}{rl}
\frac{2}{L}, \quad&kL\leq x\leq \left(k+\frac{1}{2}\right)L,\\
&\\
0,\quad&\left(k+\frac{1}{2}\right)L\leq x\leq (k+1)L,
\end{array}
\right.
\label{11}
\eeq
\noindent
where $k=0,\,1,\,2,\ldots$ This would mean that in the course of the
measurements the sphere is located with equal probability above 
different points of the
convex part of corrugation but cannot be located above the concave one.

It is even more reasonable to suppose that the function 
$\rho$ increases by the linear law when the sphere approaches 
the points of a stable
equilibrium. In this case the functional dependence is given by
\beq
\rho(x)=\left\{
\begin{array}{l}
\frac{16}{L^2}x, \quad kL\leq x\leq \left(k+\frac{1}{4}\right)L,\\
\\
\frac{16}{L^2}\left(\frac{L}{2}-x\right),\quad
\left(k+\frac{1}{4}\right)L\leq x\leq \left(k+\frac{1}{2}\right)L,\\
\\
0,\quad \left(k+\frac{1}{2}\right)L\leq x\leq (k+1)L.
\end{array}
\right.
\label{12}
\eeq

By way of example in Fig.~1 the measured Casimir force is presented 
together with experimental uncertainties acting between a corrugated
plate and a sphere \cite{21}. In the same figure the theoretical results 
computed by Eqs.~(\ref{2})--(\ref{4}) are shown by the curves 1 
(uniform distribution), 2 (distribution of Eq.~(\ref{11})), 3
(distribution of Eq.~(\ref{12})), and 4 (the bottom of the sphere 
is over the maximum of corrugation
at all times). It is seen that the curve 3 is
in agreement with experimental data in the limits of given
uncertainties $\Delta F=5\,$pN, $\Delta a=5\,$nm. The root
mean-square average deviation between theory and experiment within
the range 169.5\,nm$\leq a\leq$400\,nm (62 experimental points) where the
perturbation theory (\ref{4}) is applicable is
$\sigma=20.28\,$pN for the curve 1,
$\sigma=8.92\,$pN (curve 2),
$\sigma=4.73\,$pN (curve 3),
and $\sigma=9.17\,$pN (curve 4).
By this means perturbation theory with account of the lateral Casimir force
can be made consistent with experimental data.

In conclusion, it may be said that the lateral projection of the
Casimir force in a configuration of a sphere above a corrugated plate is
examined. It was shown that under the influence of this force 
the sphere tends 
to change its position in the direction of a nearest point of stable
equilibrium which coincides with a corrugation maximum. This effect
makes reasonable the suppositions about different probability
distributions describing location of a sphere above a corrugated plate.
A particular distribution is suggested which leads to agreement of the 
perturbation theory and experimental data representing the nontrivial
boundary dependence of the Casimir force. The final solution of the
problem may be achieved in the experiment where both the vertical
and the lateral Casimir force are measured.

\section*{Acknowledgments}
The authors are greatly indebted to U.~Mohideen for supplying them with
original experimental data and V.~M.~Mostepanenko for several helpful
discussions. G.L.K.{\ }is grateful to the Centro Brasileiro de 
Pesquisas F\'{\i}sicas where this work was performed for kind
hospitality. She was supported by FAPERJ. 

%%%%%%%%%%%%%%%%%%%%%%%%%%%%%%%%%%%%%

%%%%%%%%%%%%%%%%%%%%%%%%%%%%%%%%%%%%%
\begin{figure}[h]
\epsfxsize=15cm\centerline{\epsffile{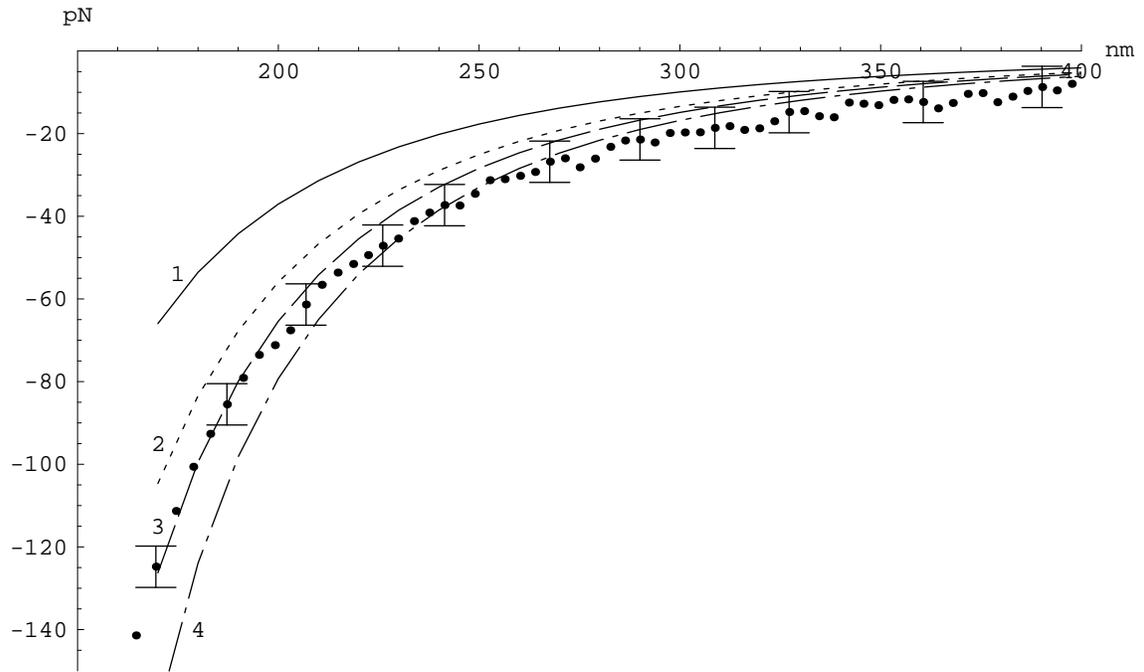} }
\vspace{0.5cm}
\caption{The Casimir force $F(a)$ from Eq.~(2)
as a function of the surface separation in
configuration of a sphere above a corrugated disk. Curve 1 represents 
the computational results obtained with the uniform probability
distribution, curve 2 --- with the distribution of Eq.~(11),
curve 3 --- of Eq.~(12), and curve 4 --- for a sphere situated above 
the points of stable equilibrium. Solid circles represent experimental
data.
}
\end{figure}
\end{document}